# Statistics of Quenched Defects Containing Semi-Flexible Polymer Chain: Exact Results (II)

**Pramod Kumar Mishra**

Department of Physics, DSB Campus, Kumaun University, Nainital (Uttarakhand) India

*Author: pkmishrabhu@gmail.com*

*Abstract*— We describe method to discuss thermodynamics of a defected semi-flexible homo-polymer chain in the two and three dimensions using fully directed self-avoiding walk lattice model. The defects are located along a line and these defects are not in the thermal equilibrium with the monomers of the semi-flexible polymer chain; i. e. we consider the case of defected semi-flexible polymer chain in the present manuscript for the case of quenched defects. There are $m$ defects on the conformations of the $N$ monomers long semi-flexible polymer chain and we exactly count the number of $Q$ realizations of the defected conformations of $N$-monomers long self-avoiding semi-flexible polymer chain; and thus we derive the exact expression of the free energy of the defected semi-flexible polymer chain for the finite length (i. e. using the fixed particle ensemble method); and we also derive exact expression of the partition function for the defected self-avoiding semi-flexible polymer chain in the thermodynamic limit using the grand canonical ensemble theory. The method described in this manuscript may be easily extended to another case of the defected polymer chain for isotropic/directed walk lattice models.

*Keywords*- Quenched defects, short chain, thermodynamic limit, conformational statistics

## I. INTRODUCTION

There are wide range of applications of the polymer materials [1],[2]. These applications involve the technology based intricacies and therefore understanding of the fundamental aspects of these polymer-macro-molecules may be useful to improve technological applications of the polymer materials. The well known applications of the polymer materials may be the steric stabilization of the dispersions of their solutions, use of the Biosensors, and coating on the degradable surfaces using polymer materials, etc. [1],[2],[3]. But the advances in the experimental techniques on the single macromolecules made it feasible to manipulate the polymer molecules; and which are due to inventions of the devices like optical tweezers, the atomic force microscope (*AFM*). The *AFM* made it possible to study conformational statistics of single macromolecules [4],[5],[6],[7],[8]. The *AFM* is widely used to measure elastic constants of single macromolecule (*DNA, Proteins*) and also to manipulate the synthetic macro-molecules for its desired applications. The conformational properties of the nano-polymer aggregates may be easily measured using the AFM.

The biological processes occurring in the living cells may be the function of the conformational changes in the Bio-molecules; and presence of the defects may affect Biological functioning of the cell and it may lead to occurrence of disease in the creature. However, on the other hand the defects may be used to synthesize polymeric materials of the desired size and it may have required conformational properties; and thus study of the defected polymer chain (i. e. the chain of desired conformational statistics) may be useful [9],[10],11],[12]. Therefore, we consider a defected single polymer chain to analyze conformational statistics of the polymer molecule for the case when the defects are not in the thermo-dynamical equilibrium with the monomers of the polymer chain; i. e. present manuscript deals with the conformational statistics of a defected semi-flexible polymer chain for the quenched defects case. However, a report on the conformational statics of the self-avoiding semi-flexible polymer under annealed defects case has been planned to be published elsewhere.

The conformational statistics of the defected self-avoiding polymer chain is least understood. Therefore, we consider the case of semi-flexible polymer chain containing quenched defects. A fully directed self-avoiding walk lattice model [13],[14] is used to mimic the conformations of the defected semi-flexible homo-polymer chain to study behavior of the self-avoiding polymer chain in the presence of quenched defects. The defects are considered to be located along a line, and the walks of the polymer chain were enumerated using a square and the cubic lattices to discuss behavior of the defected chain in the two as well as three dimensions, respectively. The conformations of the defected macromolecule may play vital role in the Biological functioning for the example; the controlled drug delivery may be possible using biodegradable matrix of the polymer-nano-aggregates [9],[10],11],[12].

The manuscript is organized as follows: In the section two, a fully directed self-avoiding walk model has been described on a square and the cubic lattices. The stiffness of the chain is accounted by assigning a Boltzmann weight of the bending energy for each bend in the chain. The defects are assumed to be located along a line (i. e. say $y=0$). The bending of the polymer chain may also be treated as the defects of the chain. We describe the methods of calculations of the thermo-dynamical parameters of the self-avoiding semi-flexible polymer chain in the section 3. We summarize method of calculating thermo-dynamical properties of the defected chain in the section 4 and conclude the manuscript in the section 5.

## II. THE MODEL AND METHOD

The lattice model of the fully directed walks [13],[14] is well known and this model is used widely to study statistics of the self-avoiding polymer chain. In this method, we consider the steps along $+x$ and $+y$ directions on a square lattice to mimic conformations or the walks of the self-avoiding polymer chain of $N$ monomers in two dimensions. While in the case cubic lattice, the steps are permitted only along $+x$, $+y$ and $+z$ directions to enumerate conformations of a polymer chain of $N$ monomers in the three dimensions. The stiffness ($k$) of the chain is included by assigning a Boltzmann weight for bending energy ($\mathcal{E}_B$) of the chain [$k=exp(-\beta\varepsilon_B)$, the term $\beta$ is the inverse temperature]; [15]. A conformation of the eleven monomers long defected semi-flexible polymer chain is shown in the figure no. 1; and the polymer chain is grafted at a point $O$; and while the defects are assumed to be located along a line (i. e. say, $y=0$). We have shown the monomers using two open circles connected through a solid line, while a pair of closed circle connected through solid line used to represent the defect, schematically. The grand canonical partition function of the defected polymer chain is obtained using methods of recursion relations [13],[14],[15].[16],[17],[18], when the self-avoiding polymer chain is containing the defects. The defect chain's contributions has been calculated separately and the general expression of the grand canonical partition function for the defected polymer chain may be written as,

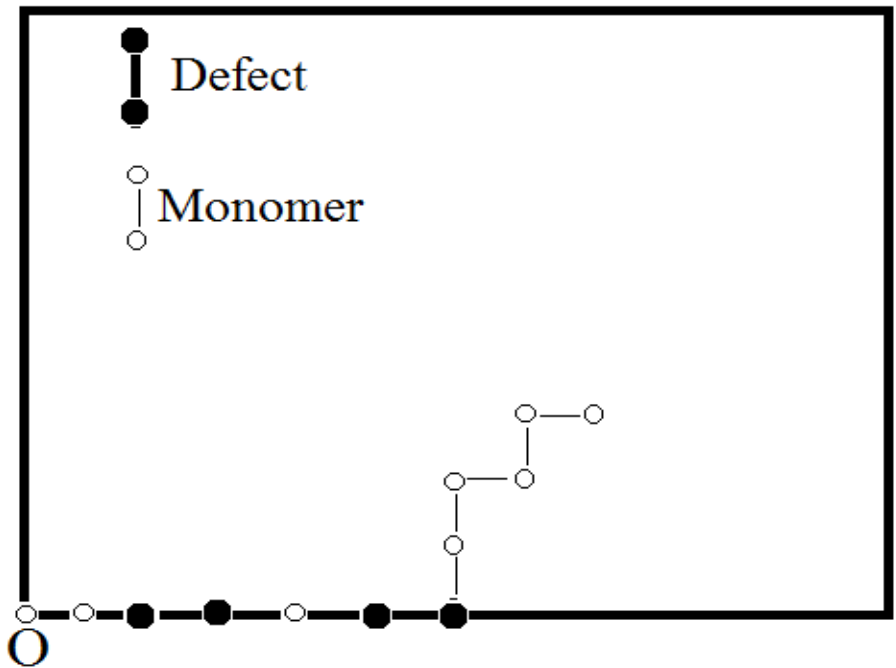

**Figure No. 1:** A defected self-avoiding chain of eleven monomers is shown in this figure schematically. One end of the polymer chain is grafted on the point $O$, and four bends are also seen in the figure. The defect is shown using a pair of solid circle connected through solid line while a pair of open circle connected through solid line represents monomer of the polymer chain.

$$G(u,g,k) = \sum_{P=1}^{N} \sum_{(N-P)=0}^{N-P} \sum_{N_B=0}^{N-1} u^P g^{N-P} k^{N_B} \qquad (1)$$

Where, $g$ is the step fugacity of the monomer of the polymer chain in the bulk, and $u\{=g*exp[exp(-\beta\varepsilon_u)]\}$ is the Boltzmann weight which corresponds to the onsite attraction-energy ($\varepsilon_u$) acting in between monomer and the defect. The term $k$ is the stiffness parameter of the semi-flexible polymer chain of $N$ monomers and but $P$ is the position of the defect along the defect line. It is possible to consider conformations of the semi-flexible polymer chain containing m number of defects in its conformations. It is possible to calculate the exact number of defected conformations of an $N$ monomers long semi-flexible polymer chain, and thus the partition function for the fixed particle ensemble and grand canonical ensembles may be obtained using following procedure. There is the maximum $d^{N-1}$ conformations of the defected polymer chain while at least $(d-1)*d^{N-1}$ conformations which are defect free, where d is the dimensionality of the space. Thus, in the case of square lattice the value of d=2 and while in the case of cubic lattice the value of d=3, and discussion is confined to the case of self-avoiding polymer chain.

### III. CALCULATIONS OF THE CONFORMATIONAL PROPERTIES

We consider fixed particle ensemble theory to discuss conformational statistics of a defected semi-flexible polymer chain. The partition function [C(P,N',k)] of the defected polymer chain under fixed particle ensemble is written as,

$$G(u,g,k) = \sum_{N=1}^{\infty} C(P, N-P, k) \tag{2}$$

Where,

$$C(P, N-P, k) = \sum_{M=R}^{N} [1 + \sum_{M=1}^{N-P-1} k^M \frac{\prod_{M=1}^{N-P-1}(N-M)}{(M-1)!}] \tag{3}$$

There are $P$ monomers along the defect line and remaining $(N-P)$ monomers are located in the bulk, and thus; the number of defected conformations of the polymer chain containing $N_B$ bends may be written as,

$$C(R,N,k) = \sum_{P=R}^{N} [1 + \sum_{M=1}^{N-P-1} (d-1)^M k^M \frac{\prod_{M=1}^{N-P-1}(N-M)}{(M-1)!}] \tag{4}$$

$$C(R,N,k) = \sum_{P=R}^{N} [1 + \sum_{M=1}^{N-P-1} (d-1)^M k^M C_M^{N-M-1}] \tag{5}$$

Where,$\{C(P,N,N_B=0)=1\}$; and d is the dimensionality of the space and its value is 2, 3 for the square and the cubic lattices; respectively.
Since, the monomers of the polymer chain and the defects are not in the thermal equilibrium and therefore, the Helmholtz free energy $F_Q(u, g, k)$ of the defected semi-flexible polymer chain for the present case of the quenched defects may be written as,

$$F_Q(u,g,k) = \frac{1}{Q}\{-k_B T * \sum_{P=1}^{Q} Log[C(R, N, k)]\} \tag{6}$$

The symbol $Q$ is used for the number of realizations of the conformations containing the quenched defects and it is equal to C(R, N, k=1).
The thermo-dynamical properties of the defected chain may be calculated using above equation nos. (2) and (3) provided the length of the polymer chain is larger in comparison to number of realizations ($Q$) of the $m$ defects distributions; and it is merely because one end of the chain is grafted at a point $O$. The method described in the present report may be extended to other cases of self-avoiding polymer chain in the disordered environment and therefore it is useful approach to control thermo-dynamical statistics of the defected chain.

It is also to be noted that once the free energy of the defected chain is known other thermo-dynamical parameters of the defected chain may follow easily from the Helmholtz free energy or following equations:
The free energy of quenched defected $N$-monomers long semi-flexible polymer chain may be written as,

$$\frac{F_Q(R,N,k)}{k_B T'} = Log(g)^N - Log[f(k)] \qquad (7)$$

Where,

$$T' = \frac{T}{Q}, \qquad g^{-1} = \sqrt{\frac{C(R, N+2, k=1)}{C(R, N, k=1)}}, \qquad f(k) = \frac{C(R, N, k)}{C(R, N, k=1)} \qquad (8)$$

Using above quoted relations we may find average length of the defected semi-flexible polymer chin and also the number of the bends in the polymer chain to obtain persistence length of the chain.

$$<N> = -\frac{\partial Log(\frac{F}{k_B T'})}{\partial Log(g)} \qquad (9)$$

$$<N_B> = -\frac{\partial Log(\frac{F}{k_B T'})}{\partial Log(k)} \qquad (10)$$

We define persistence length ($l_P$) as the average length of the chain in between its two successive bends [15],

$$<l_P> = \frac{<N>}{<N_B>} \qquad (11)$$

## IV. THE SUMMARY AND DISCUSSIONS

We described method to obtain the equilibrium statistics of a defected self-avoiding semi-flexible polymer chain using fully directed self-avoiding walk lattice model in two and three dimensions. The defects are located along a line and the monomers of the chain and the defects are not in the thermal equilibrium. Therefore, we report thermodynamics of the defected self-avoiding semi-flexible polymer chain under quenched defects case. There may be m defects located along the line (i. e. say y=0), and Q conformations of the self-avoiding polymer chain were considered and all these conformations are said to contain m defects in its segments (conformations).

We obtained analytical expression of the Helmholtz free energy for the self-avoiding defected semi-flexible polymer chain. The expression of the Helmholtz free energy may be used to calculate other thermo-dynamical parameters of the defected semi-flexible polymer chain.

The method described in the present manuscript may be extended to other cases of self-avoiding semi-flexible polymer chain and also for other distributions of the defects in the two and three dimensions.

## V. CONCLUSIONS

The statistics of the self avoiding polymer chain were well under stood and there is no report available in the literature regarding conformational statistics of a self-avoiding polymer chain containing quenched defects. Therefore, we consider a self-avoiding linear semi-flexible homo-polymer chain in two and three dimensions and the quenched defects are assumed to be located along a line. The analytical expression were obtained regarding partition function for the short chain and also for the chain length in the thermo-dynamical limit and thus accordingly the expression of the Helmholtz free energy of the polymer chain containing m defects and Q conformations (i. e. realizations) of the defects were taken in to consideration to describe conformational statistics of the quenched defected semi-flexible polymer chain. The proposed method may be extended to another case of defects distribution and other versions of the directed/isotropic walk models for the self-avoiding semi-flexible polymer chain.

The methods described above regarding calculations of the thermo-dynamical parameters for defected chain may be useful to calculate the statistics of the partially directed self-avoiding semi-flexible polymer chain where the details of the defected conformations is exactly known. It is also possible to calculate size of the defected polymer chain along and perpendicular to the line of the defects using above described methods of the calculations of the thermo-dynamical parameters of the semi-flexible polymer chain.

## ACKNOWLEDGMENTS

Author would like to thank his mentor Professor Yashwant Singh, Institute of Science, Banaras Hindu University, Varanasi (U. P.) India, for the training.

## AUTHOR'S PROFILE

**Author: Pramod Kumar Mishra** (also know as **P. K. Mishra**);
Pramod Kumar Mishra is serving to the Kumaun University since 19.03.2005 and he obtained B. Sc., M. Sc. and PhD degrees from Banaras Hindu University (Varanasi) Uttar Pradesh, INDIA. He cleared the National Eligibility Test in the year 1998 i. e. at the time of passing or completing the M. Sc. degree in the Solid State Physics. The area of his research interest is theoretical modeling of the polymer systems (macro-molecules) using the statistical Physics and the computational methods. He published several articles as a sole author in the journal of international repute. He also published a few popular artciles on the science and the humanities related topics.